\documentclass[11pt]{article}

 \usepackage{epsfig,moriphoto,amssymb}

\def\Journal#1#2#3#4{{#1} {\bf #2}, #3 (#4)}

\def\be{\begin{equation}}
\def\ee{\end{equation}}
\def\bea{\begin{eqnarray}}
\def\eea{\end{eqnarray}}

\begin{document}
\vspace*{4cm}
\title{ON A QCD-BASED PION DISTRIBUTION AMPLITUDE VS. RECENT EXPERIMENTAL
DATA}

\author{A.P. BAKULEV${}^{1,\ }$\footnote{talk presented by this author},
        S.V. MIKHAILOV${}^{1}$, N.G. STEFANIS${}^{2}$}

\address{${}^{1}$Bogoliubov Laboratory of Theoretical Physics,
         JINR, 141980 Dubna, Russia  \\
         ${}^{2}$Institut f\"ur Theoretische Physik II,
         Ruhr-Universit\"at Bochum,
         D-44780 Bochum, Germany}
\maketitle
\abstracts{%
 Using QCD sum rules with nonlocal condensates
 the twist-2 pion distribution amplitude is determined
 by means of its moments and their confidence intervals,
 including also radiative corrections.
 An admissible set of pion distribution amplitudes is constructed
 in the $a_{2}$, $a_{4}$ plane of the Gegenbauer polynomial
 expansion coefficients.
 The determined $a_{2}$, $a_{4}$ region strongly overlaps
 with that extracted from the CLEO data by Schmedding and Yakovlev.
 Comparisons are given with results from Fermilab experiment E791
 and recent lattice calculations.
}

\section{Pion Distribution Amplitude}
The main object of this talk is the pion distribution amplitude
(DA), defined by
\be
 \langle 0\mid \bar d(z)\gamma_{\mu}\gamma_5 E(z,0)
             u(0)\mid \pi(P) \rangle \Big|_{z^2=0}
 = i f_{\pi}P_{\mu}
    \int_{0}^{1} dx\ e^{ix(zP)}\
     \varphi_{\pi}(x,\mu^2) \; ,
\ee
where gauge invariance is ensured by the Fock-Schwinger string
$E(z,0) = {\cal P}\exp\left[i g \int_0^z A_\mu(t) dt^\mu\right]$.

The pion DA has the following important properties:
(1) it is multiplicatively renormalizable,
(2) it has isospin symmetry:
 $\varphi_{\pi}(1-x,\mu^2) = \varphi_{\pi}(x,\mu^2)$,
 and
(3) its normalization is conserved:
 $\int_{0}^{1} dx\ \varphi_{\pi}(x,\mu^2) = 1$.

 Pion DAs naturally appear in the perturbative QCD (pQCD)
description of any {\it hard exclusive} process with pions.
For example, the form factor of $\gamma^*\gamma^*\to\pi^0$
decay with $-q_{1,2}^2\sim Q^2 \geq 1$~GeV$^2$
is factorized in pQCD according to
\be
 F_{\gamma^*\gamma^*\to\pi^0}(q_1^2,q_2^2) =
 C(q_1^2,q_2^2;\mu^2;x) \otimes
 \varphi_\pi(x;\mu^2) + O\left(Q^{-4}\right) + \ldots
\ee

For $\varphi _{\pi}(x)$ we use an expansion in terms of
eigenfunctions of the 1-loop evolution kernel,
$x\bar{x}C^{3/2}_n(2x-1)$,
\be
 \varphi_\pi(x;\mu^2) = \varphi_\pi^{\mbox{\rm \footnotesize as}}(x)
  \Bigl[ 1
       + a_2(\mu^2) C^{3/2}_2(\xi)
       + a_4(\mu^2) C^{3/2}_4(\xi) + \ldots \Bigr]
\label{eq:DA_Rep}
\ee
with $\varphi_\pi^{\mbox{\rm \footnotesize as}}(x)\equiv 6 x \bar{x}$
being the asymptotic pion DA and $\xi \equiv 2x-1$.
In this expansion all scale-dependence
is accumulated in the coefficients
$\left\{a_2(\mu^2), a_4(\mu^2), \ldots \right\}$.
Note that the evolution of the pion DA at the 2-loop level
is available from~\cite{MR86ev,Mul94}.

How one can obtain the $\varphi_{\pi}(x,\mu^2)$?
It is possible to extract it from:\\
$\bullet$~~experimental data: (i) see the recent papers
  of the CLEO Collaboration~\cite{CLEO98}
  and the analysis of Schmedding and Yakovlev of these
  data~\cite{SchmYa99},
  and (ii) using the data of the E791 Collaboration~\cite{Ash00}\\
$\bullet$~~QCD Sum Rules with Non-Local Condensates (NLC)
  see~\cite{MR86,BM95,BMS01}\\
$\bullet$~~transverse lattice simulations \cite{Dal01,BuSe01}\\
$\bullet$~~instanton-induced models \cite{ADT00,PPRWG99}.\\
In this talk we consider all these sources separately, but the main
focus is on the first 2 items.

\section{Revision of the NLC QCD SR Results}
We re-analyze our NLC SRs with the modification of
one of the antiquark-gluon-quark NLC contributions to
obtain revised values of the moments
 $\langle{\xi^N}\rangle_\pi = \int_0^1 \varphi_\pi(x;\mu^2)
  \left(2x-1\right)^N dx$
  ($N=2,\ldots,10$),
new estimates of error-bars, and a new estimate of
 $\langle{x^{-1}}\rangle_{SR}
 =\int_0^1 \varphi_\pi(x;\mu^2)x^{-1} dx $,
 where a SR is used that is constructed directly for this quantity.

Our model of NLCs, illustrated by the scalar NLC
$
 \langle{\bar{q(0)}q(x)}\rangle
 = \langle{\bar{q(0)}q(0)}\rangle \exp\left(-|x^2|\lambda_q^2/8\right)$,
uses only a single parameter $\lambda_q^2$, which is related to
the vacuum (fields) correlation length $1/\lambda_q=\Lambda$ the
latter being of the order of the hadron size, as estimated in
non-pQCD approaches:
\be
 \lambda_q^2=\left\{
 \begin{array}{lc}
 0.4\pm0.1~\mbox{GeV}^2 &
  \left[\mbox{\ QCD SRs, 1983~\cite{BI82} }  \right]\\
 0.5\pm0.05~\mbox{GeV}^2 &
  \left[\mbox{\ QCD SRs, 1991~\cite{Piv91}\ }  \right]\\
 \approx 0.5~\mbox{GeV}^2 &
  \left[\mbox{\ Lattice, 1998-99~\cite{DDM99,Meg99}\ } \right]
 \end{array}
 \right.
\ee

From the lhs of Fig.~1, one sees that the quality of the stability
in the NLC QCD SRs for $\lambda_q^2 = 0.4$~GeV$^2$ is quite high
(solid line stands for the best threshold $s_0=2.2$~GeV$^2$,
dashed lines -- for 10\%-variations of this parameter).
\begin{figure}[thb]
$$\psfig{figure=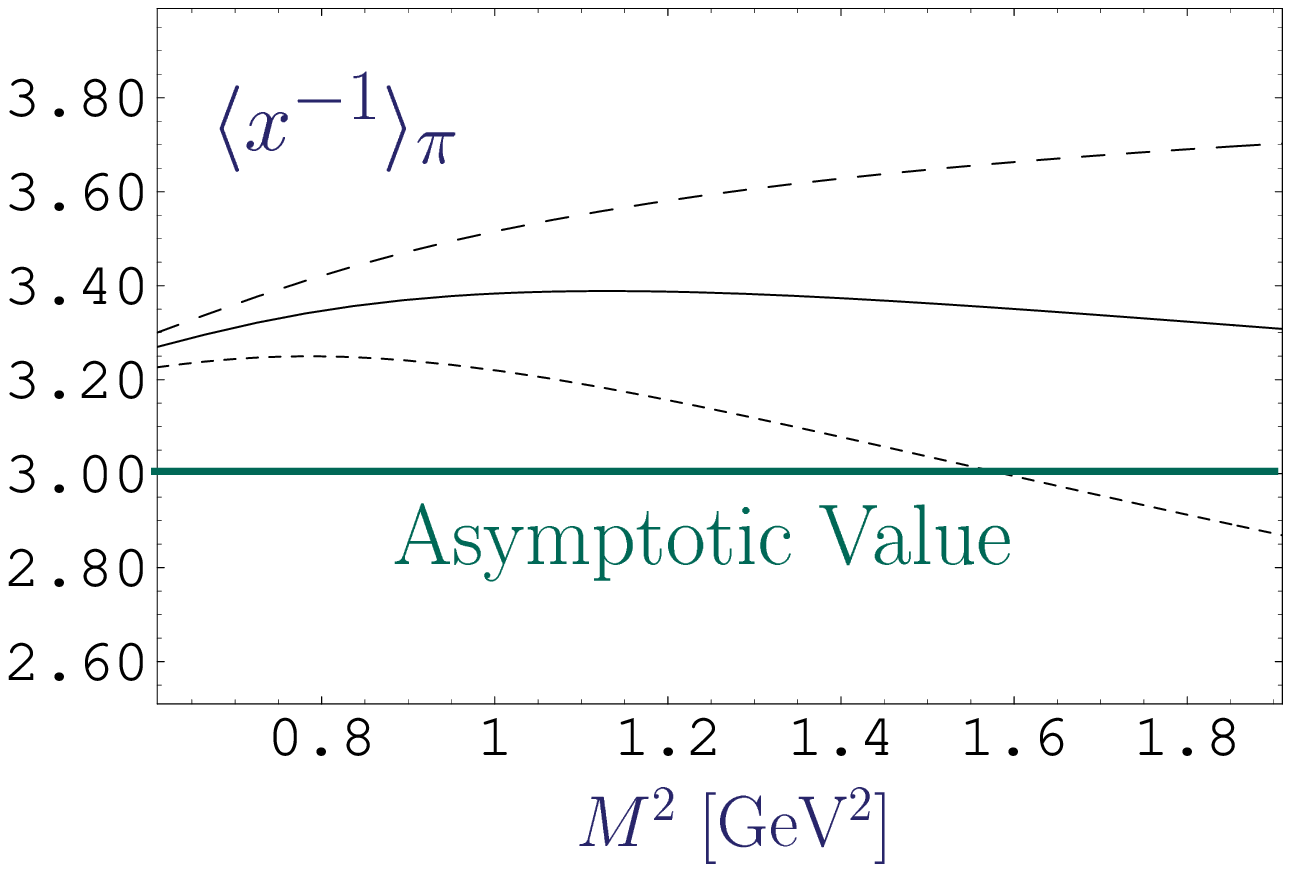,height=30mm}
   ~~~~~~
   \psfig{figure=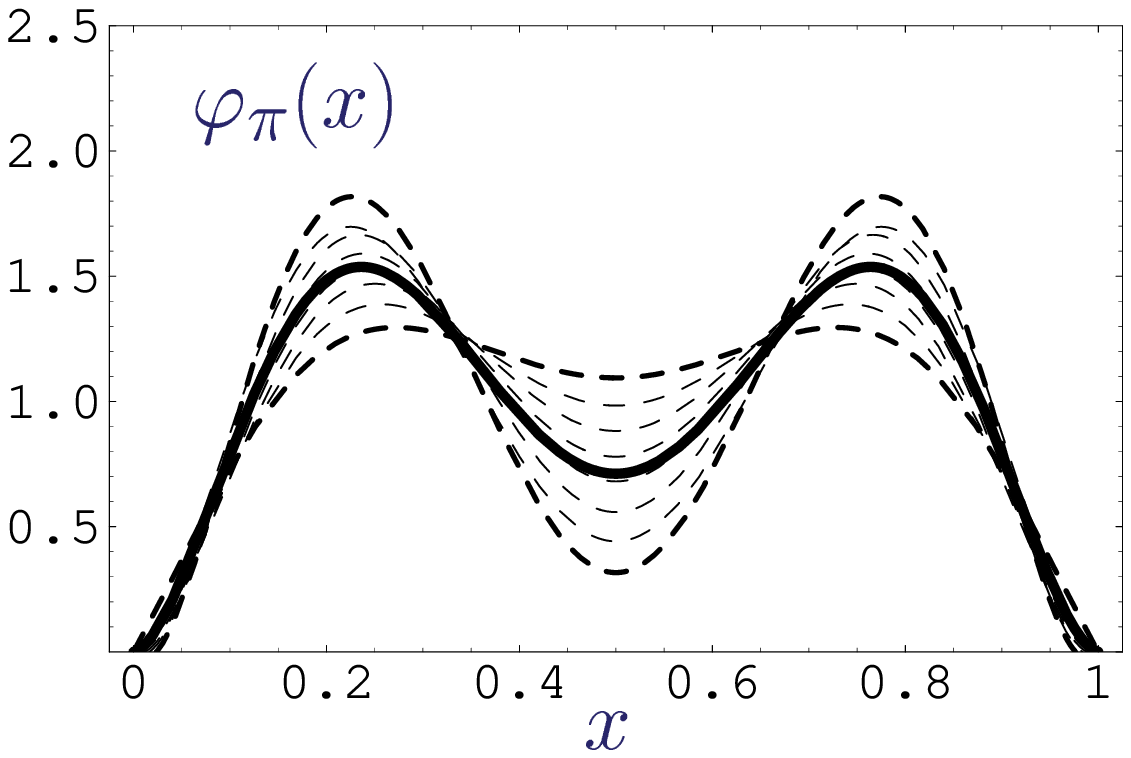,height=30mm}
   ~~~~~~
   \psfig{figure=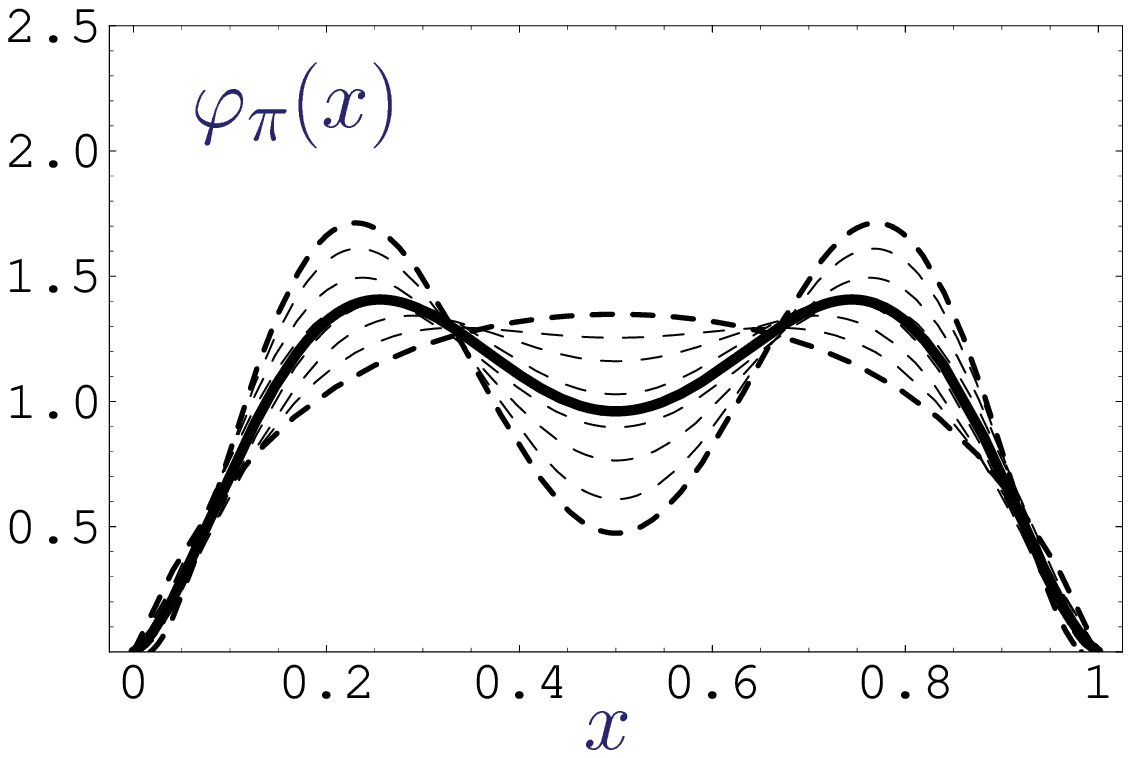,height=30mm}
    $$
    \vspace*{-10mm}

    \caption{Left side: stability curves for $\langle x^{-1}\rangle_\pi$
     against the Borel parameter $M^2$.
     Central part: Bunch of admissible DAs corresponding to
     $\lambda^2_{q}=0.4$~GeV$^2$ with best-fit parameters
     $a_2 = + 0.188$, $a_4 = - 0.130$.
     Right side: Same as central part, but for
     $\lambda^2_{q}=0.5$~GeV$^2$ with best-fit parameters
     $a_2 = + 0.126$, $a_4 = - 0.091$.}
\end{figure}
The obtained moments are shown on the lhs of Fig.~2. We
reconstruct the pion DA from these five moments, using models
(\ref{eq:DA_Rep}) at $\mu^2=1$~GeV$^2$ with two non-zero
Gegenbauer coefficients. The best-fit DAs obtained this way (with
$\chi^2\approx10^{-3}$) are shown in Fig.~1 as thick solid lines.
The corresponding error bars to the DAs, allowed by the moment
SRs, are also shown. The bunches of these broken lines represent
the self-consistent DAs in the sense that the value of the
associated inverse moment, $\langle x^{-1}\rangle_\pi = 3.17(8)$
($\langle x^{-1}\rangle_\pi  = 3.13(8)$, for
$\lambda_q^{2}=0.5$~GeV${}^{2}$), is in good agreement with the
value determined from the special SR:
$\langle
x^{-1}\rangle_\pi^{\mbox{\footnotesize SR}} = 3.33(32)$ ($\langle
x^{-1}\rangle_\pi^{\mbox{\footnotesize SR}} = 3.19(32)$).

\section{Comparing with CLEO results in SY approach}
Schmedding and Yakovlev~\cite{SchmYa99} have provided a useful
analysis of the CLEO data on the $\gamma^*(q)\gamma\to\pi^0$ form
factor~\cite{CLEO98}, using light-cone QCD SRs and taking into
account NLO and twist-4 corrections. They estimated the first two
Gegenbauer coefficients of the pion DA and obtained \be
 a_2 = 0.19
     \pm 0.04(\mbox{stat})
     \pm 0.09(\mbox{sys})\; , \quad
 a_4 = -0.14
      \pm 0.03(\mbox{stat})
      \mp 0.09(\mbox{sys})
\ee
with results displayed in Fig.~2 in the form of confidence regions
in terms of $a_2$ and $a_4$.

We evolve our allowed sets to the CLEO scale $\mu^2=(2.4~\mbox{GeV})^2$
and insert them directly into the SY plot to obtain results in the
$a_2, a_4$ plane, shown in Fig.~2.
\begin{figure}[thb]\noindent
 $$\psfig{figure=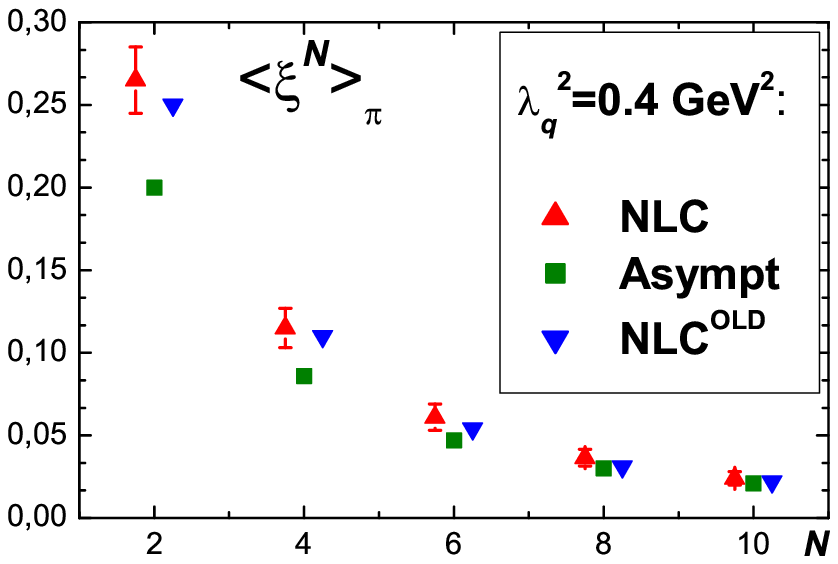,height=30mm}
   ~~~~~~
   \psfig{figure=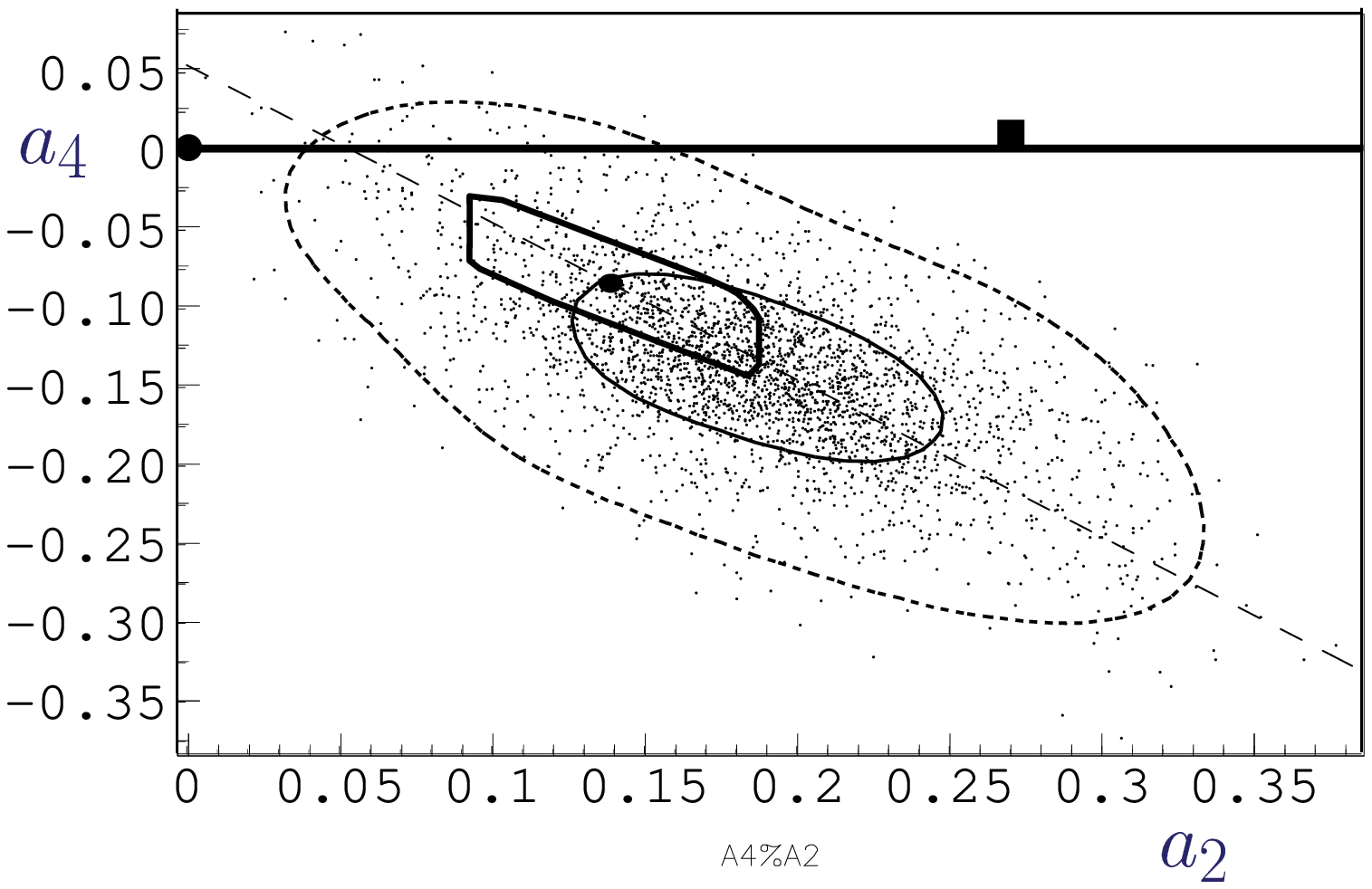,height=30mm}
   ~~~~~~
   \psfig{figure=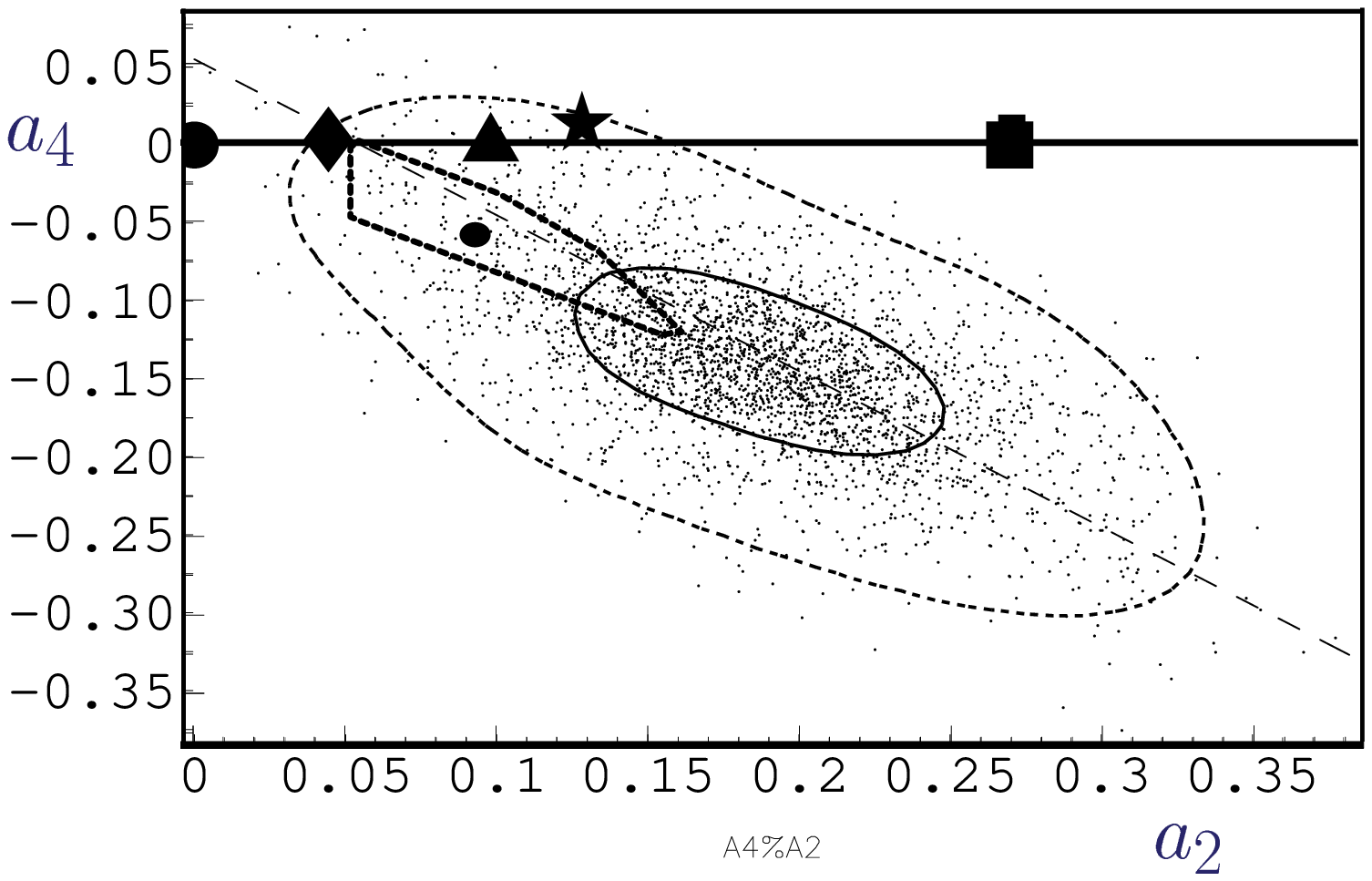,height=30mm}
    $$
    \vspace*{-10mm}

    \caption{Left: $\langle \xi ^{N} \rangle$ against $N$.
     Central part: Confidence region of pion DAs for
     $\lambda^2_{q}=0.4$~GeV$^2$.
     Right side: the same as before with specific explanations given
     in the text, but with
     $\lambda^2_{q}=0.5$~GeV$^2$.}
\end{figure}
Inspection of these plots reveals that bunch-1
$\left[\lambda_q^2 = 0.4~\mbox{GeV}^2\right]$
and bunch-2
$\left[\lambda_q^2 = 0.5~\mbox{GeV}^2\right]$
are intersecting with the SY $1\sigma$-region
and are thus quite {\em compatible} with CLEO data.

Inspecting the SY plot, one natural question arises: why the
confidence regions are stretched along the diagonal $a_2 + a_4 =
{\rm const}$~? To answer this question, it is instructive to
analyze the CLEO data, employing pure pQCD: \be \label{eq:FFpQCD}
 \frac{3Q^2}{4\pi} F_{\gamma\gamma^*\pi}
 \approx Q^2\cdot C \otimes \varphi_\pi
 \ =\ \langle{x^{-1}}\rangle_\pi
   + 3\alpha_s\left(\Delta_0+\Delta_2+\Delta_4\right)
   + \ldots
\ee
We see, that, up to radiative corrections, CLEO in fact
measured the inverse moment of the pion DA,
$\langle{x^{-1}}\rangle_\pi$, which is simply connected to the
diagonal combination of the $a_2$ and $a_4$ coefficients:
\be
 \langle{x^{-1}}\rangle_\pi = 3\left(1 + a_2+a_4\right)
\ee
The SY analysis gives: $a_2+a_4 = 0.05 \pm 0.07$, whereas the
special NLC QCD SRs yields for this moment
$\frac13\langle{x^{-1}}\rangle_\pi^{\mbox{\footnotesize SR}}-1 =
0.10 \pm 0.10$ (at $\mu = 1$~GeV), and our bunch-1 of the allowed
pion DAs produces ${a_2+a_4} = 0.056 \pm 0.03$ (at $\mu =
2.4$~GeV), in excellent agreement with S\&Y result.

It is useful to have a look on the numerical values of different
terms in~(\ref{eq:FFpQCD})
\bea
 \frac{3Q^2}{4\pi} F_{\gamma\gamma^*\pi}
 &\approx& 3\left[
            1 + \left(a_2 + a_4\right)
              + \alpha_s\Delta_0
              + \alpha_s\left(\Delta_2 + \Delta_4\right)
              + \ldots\
            \right]\nonumber\\
   & = &   3\left[
            1 + \ \ \ 0.05
      \ \ \ \ - \ 0.17
          \ \ - \ \ \ 0.014
        \ \ \ + \ \ \ldots\
            \right]\; .
\eea
The CLEO data gives numerically
$\left(3Q^2/4\pi\right)F_{\gamma\gamma^*\pi} \approx 2.45$. We see
that in the LO pQCD analysis (i.e., when all $\alpha_s\Delta_N =
0$), one arrives at the estimate:
$\langle{x^{-1}}\rangle_\pi=2.45$. From this, one might conclude
that $\varphi_\pi(x)$ should be \textit{narrower} than the
asymptotic one~\cite{RR96}. Taking into account only the main part
of the NLO correction ($\alpha_s\Delta_0 = -0.17$), we get the
estimate, $\langle{x^{-1}}\rangle_\pi=2.96$, and conclude that
$\varphi_\pi(x)$ could have the same width as the asymptotic DA.
But the full NLO (plus twist-4 contribution) light-cone QCD SR~\cite{SchmYa99} 
provides instead, 
$\langle{x^{-1}}\rangle_\pi \approx 3.15$, indicating that
\textit{$\varphi_\pi(x)$ should be \textit{broader} than the
asymptotic DA}~\cite{SSK00}, just as it appears in our bunches in
Fig.~1.

The E791 collaboration has measured dijet production in
diffractive $\pi A$ interactions~\cite{Ash00}. Such an experiment
has been suggested in 1993 in~\cite{FMS93} as a means of measuring
\textit{directly} the squared pion DA at large transverse momentum
transfers. We obtain a good fit of the E791 data using our model
(symbol $\star$ on the rhs of Fig.~2) with $a_{2}^{\mbox{\rm
\footnotesize fit}}=0.12$ and $a_{4}^{\mbox{\rm \footnotesize
fit}}=0.01$ at the scale $\sim 8$~GeV$^2$. The resulting value of
the ``diagonal'' at the CLEO scale appears to be too large:
$a_2^{\mbox{\rm \footnotesize fit}}
+a_4^{\mbox{\rm \footnotesize fit}} \simeq 0.14$. 
In our view, the interpretation of this
experiment is still questionable.
Moreover, it seems that errors are too large and should be
estimated more carefully.

There are two recent papers involving transverse lattice
simulations. Dalley~\cite{Dal01} produced (see symbol
$\blacktriangle$ on the rhs of Fig.~2)
\be
 \varphi_\pi^{\mbox{\rm \footnotesize lat}}(x;\mu^2\simeq 1~\mbox{GeV}^2)
     = 6 x \bar{x}
        \Bigl[ 1 + 0.133 C^{3/2}_2(2x-1)
        \Bigr]\; ,
\ee
and, on the other hand, Burkardt and Seal~\cite{BuSe01} arrived --
using the same approach -- at a different DA (denoted by $\bullet$
in Fig.~2) very close to the asymptotic pion DA. Note that this
large difference seems to indicate that the errors of this method
are still large and should be estimated more precisely.

The existing predictions from instanton-induced
models~\cite{ADT00,PPRWG99} are too close to $\varphi _{\pi}^{\rm
as}$, except for the new model by Prasza{\l}owicz and
Rostworowski~\cite{Pra01}, which is just outside the confidence
region of the $\varphi_2$-bunch and on the boundary of the
$95\%$-region of SY (symbol $\blacklozenge$ on the rhs of Fig.~2).

\section*{Acknowledgments}
This work was supported in part by the Russian Foundation for
Fundamental Research (contract 00-02-16696), INTAS-CALL 2000 N 587,
the Heisenberg--Landau Program (grants 2000-15 and 2001-11), and
the COSY Forschungsprojekt J\"ulich/Bochum.

\end{document}